\documentclass[twocolumn,citeautoscript,showpacs,preprintnumbers,amsmath,amssymb,prb,floatfix,footinbib]{revtex4}
\usepackage{graphicx}
\usepackage{dcolumn}
\usepackage{bm}
\usepackage{times}
\begin{document}
\title{Magnetic order in GdBiPt studied by x-ray resonant magnetic scattering}
\author{A.~Kreyssig$^1$}
\author{M.\,G.~Kim$^1$}
\author{J.\,W.~Kim$^2$}
\author{S.\,M.~Sauerbrei$^1$}
\author{S.\,D.~March$^1$}
\author{G.\,R.~Tesdall$^1$}
\author{S.\,L.~Bud'ko$^1$}
\author{P.\,C.~Canfield$^1$}
\author{R.\,J.~McQueeney$^1$}
\author{A.\,I.~Goldman$^1$} 
\affiliation{\\\textsuperscript{1}Ames Laboratory, U.\,S. DOE and 
Department of Physics and Astronomy, Iowa State University, Ames, 
Iowa 50011, USA} \affiliation{\\\textsuperscript{2}Advanced 
Photon Source, Argonne National Laboratory, Argonne, Illinois 
60439, USA}

\date{\today}

\begin{abstract}
Rare earth ($R$) half-Heusler compounds, $R$BiPt, exhibit a wide 
spectrum of novel ground states.  Recently, GdBiPt has been 
proposed as a potential antiferromagnetic topological insulator 
(AFTI). We have employed x-ray resonant magnetic scattering to 
elucidate the microscopic details of the magnetic structure in 
GdBiPt below $T_N$~=~8.5\,K.  Experiments at the Gd $L_2$ 
absorption edge show that the Gd moments order in an 
antiferromagnetic stacking along the cubic diagonal [1\,1\,1] 
direction satisfying the requirement for an AFTI, where both 
time-reversal symmetry and lattice translational symmetry are 
broken, but their product is conserved.
\end{abstract}

\pacs{75.25.-j, 75.50.Ee, 73.20.-r}


\maketitle

The discovery of three-dimensional topological-insulating states
in binary alloys (Bi$_{1-x}$Sb$_x$)\cite{Fu_2007, Hsieh_2008} and
compounds (Bi$_2$Se$_3$, Bi$_2$Te$_3$,
Sb$_2$Te$_3$)\cite{Xia_2009, Zhang_2009, Chen_2009}, which 
feature an insulating gap in the bulk but with topologically 
protected conducting states on the surfaces or edges, has opened 
a new frontier for fundamental condensed matter physics 
research.\cite{Hasan_and_Kane}  As pointed out in several papers, 
the novel properties of this class of materials offer potential 
for technological breakthroughs in quantum computing and 
magneto-electronic applications.\cite{Hasan_and_Kane, Moore_2010, 
Li_2010}  Over the past year, attention has turned towards 
investigations of new phenomena that arise when topological 
insulators (TI) also manifest, or are in close proximity to, 
other phenomena including magnetic order and 
superconductivity.\cite{Li_2010, Hosur_2010, Mong_2010, 
Hasan_and_Kane}  Recently, the Heusler and half-Heusler compounds 
have been subject to intense scrutiny because of their potential 
as TI with tunable electronic properties.\cite{Chadov_2010, 
Lin_2010, Xiao_2010, Li_2011} Specifically, Mong \emph{et 
al.}\cite{Mong_2010} have proposed that GdBiPt may provide the 
first realization of an antiferromagnetic topological insulator 
(AFTI), where both time-reversal symmetry and lattice 
translational symmetry are broken, but their product is 
conserved. Predictions for this class of TI include gapped states 
on some surfaces, gapless states on others, and novel 
one-dimensional metallic states along step edges on the gapped 
surfaces.\cite{Mong_2010}

More generally, Heusler and half-Heusler compounds exhibit a wide 
spectrum of novel ground states.\cite{Graf_2011} The rare earth 
($R$) half-Heusler compounds, $R$BiPt, feature magnetic ordering 
(GdBiPt),\cite{Canfield_1991} superconductivity (LaBiPt, 
YBiPt)\cite{Goll_2008, Butch_2011} and heavy-fermion behavior 
(YbBiPt)\cite{Fisk_1991} .  Although the low-temperature ground 
states of the $R$BiPt system (for $R$ = Ce, Nd, Sm, Gd, Tb, Dy, 
Ho, Er, Tm, and Yb) have been characterized as antiferromagnetic 
through thermodynamic and transport measurements, there have been 
few magnetic structure determinations for this 
series.\cite{Wosnitza_2006}  GdBiPt has the highest $T_N$ of the 
series at approximately 8.5\,K\cite{Canfield_1991} and, since the 
orbital angular moment $L$~=~0 for the $S$-state Gd ion, the 
magnetic structure in the absence of crystalline electric field 
effects may be directly investigated. However, the high 
neutron-absorption cross section for naturally occurring Gd is 
problematic for conventional magnetic diffraction experiments.

\begin{figure}
\centering\includegraphics[width=0.75\linewidth]{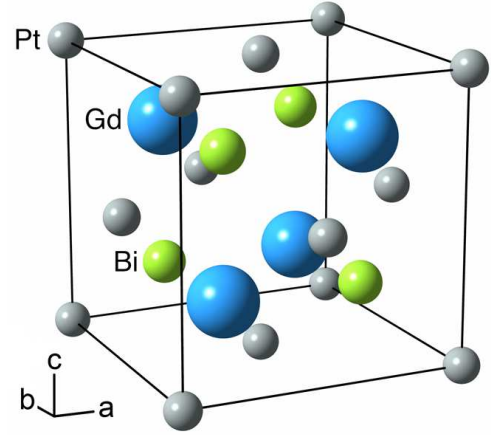}\\
\caption{(Color online) Crystal structure of GdBiPt.} \label{fig1}
\end{figure}

Here we describe the magnetic order of GdBiPt below $T_N$~=~8.5\,K
determined by x-ray resonant magnetic scattering (XRMS) at the Gd
$L_2$ absorption edge.  GdBiPt crystallizes in the MgAgAs-type
structure (cubic space group $F\,\overline{4}\,3\,m$,
$a$~=~6.68\,${\AA}$ with Gd, Bi and Pt at the 4$c$, 4$d$ and 4$a$
sites, respectively; see Fig.~\ref{fig1}).\cite{Dwight_1974,
Robinson_1994} The structure may be viewed as three sets of
elementally pure, interpenetrating face-centered cubic lattices. 
We find that the commensurate magnetic order doubles the cubic 
unit cell along the diagonal [1\,1\,1] direction, characterized 
by a propagation vector 
$\textbf{q}_m$~=~($\frac{1}{2}\,\frac{1}{2}\,\frac{1}{2}$), so 
that alternating ferromagnetic (1\,1\,1) planes of Gd are 
antiferromagnetically coupled along the [1\,1\,1] direction. This 
structure is quite similar to the model B magnetic structure for 
an AFTI via spin-orbit coupling as described by Mong \emph{et 
al.},\cite{Mong_2010} but we find that the moment direction in 
GdBiPt is not parallel to the magnetic propagation vector as is 
found, for example, in MnSbCu\cite{Forster_1968} or 
CeBiPt.\cite{Wosnitza_2006}

Single crystals of GdBiPt were solution-grown using a Bi flux and 
emerged with sizeable facets perpendicular to the [0\,0\,1] 
direction and smaller facets perpendicular to [1\,1\,1]. 
High-purity Gd (obtained from Ames Laboratory), Pt, and Bi were 
placed in an alumina crucible in the ratio 
Gd\,:\,Pt\,:\,Bi\,=\,3\,:\,3\,:\,94, sealed in a silica ampule, 
and slowly cooled from 1170$^{\circ}$C to 600$^{\circ}$C over 200 
hours. At 600$^{\circ}$C, the excess Bi solution was decanted 
from the GdBiPt crystals.\cite{Canfield_1992}  The dimensions of 
the single crystal studied in the XRMS measurements were 
approximately 3$\times$3$\times$2\,mm$^3$ with a large as-grown 
facet perpendicular to [0\,0\,1].  The measured mosaicity of the 
crystal was less than 0.01\,degrees full-width-at-half-maximum 
(FWHM), attesting to the high quality of the sample. The XRMS 
experiment was performed on the 6ID-B beamline at the Advanced 
Photon Source at the Gd $L_2$-edge ($E$~=~7.934\,keV). The 
incident radiation was linearly polarized perpendicular to the 
vertical scattering plane ($\sigma$-polarized) with a beam size 
of 0.5\,mm (horizontal) $\times$~0.2\,mm (vertical). In this 
configuration, dipole resonant magnetic scattering rotates the 
plane of linear polarization into the scattering plane 
($\pi$-polarization). For some of the measurements, pyrolytic 
graphite PG (0\,0\,6) was used as a polarization analyzer to 
suppress the charge and fluorescence background relative to the 
magnetic scattering signal. For measurements of the magnetic 
reflections, the sample was mounted at the end of the cold finger 
of a closed-cycle cryogenic refrigerator with the ($H H L$) plane 
coincident with the scattering plane.

\begin{figure}
\centering\includegraphics[width=0.88\linewidth]{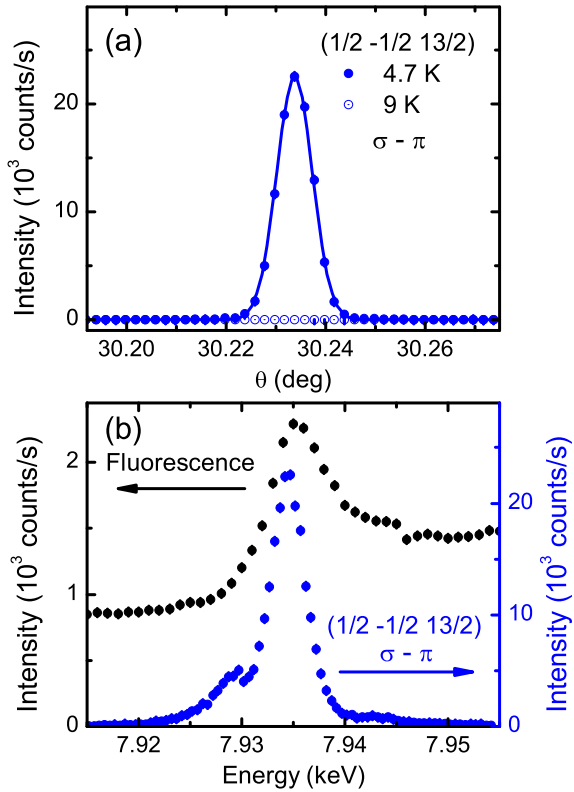}\\
\caption{(Color online) Resonant magnetic scattering from the GdBiPt
single crystal.  (a) Rocking scans ($\theta$) through the
($\frac{1}{2}$~-$\frac{1}{2}$~$\frac{13}{2}$) magnetic peak position
above (open circles) and below (filled circles) $T_N$ taken in
$\sigma$-$\pi$ scattering geometry.  (b) Energy scan through the Gd
$L_2$ absorption edge at the
($\frac{1}{2}$~-$\frac{1}{2}$~$\frac{13}{2}$) magnetic peak position
at $T$~=~4.7\,K (blue filled circles) along with the measured x-ray
fluorescence from the sample (black filled circles).} \label{fig2}
\end{figure}

Measurements of the diffraction from the sample performed in the
$\sigma$-$\pi$ scattering geometry using the PG (0\,0\,6)
polarization analyzer are shown in Fig.~\ref{fig2}. For
temperatures above $T_N$ = 8.5\,K, only Bragg peaks consistent
with the chemical structure\cite{Dwight_1974, Robinson_1994} of
GdBiPt were observed. However, upon cooling below $T_N$,
additional Bragg scattering at half-integer values of ($H K L$)
was found as shown in Fig.~\ref{fig2}(a).  The magnetic origin of
these peaks was confirmed by energy scans through the Gd $L_2$
absorption edge and from the temperature dependence of the
diffraction peak intensity as described below.  

\begin{figure}
\centering\includegraphics[width=0.88\linewidth]{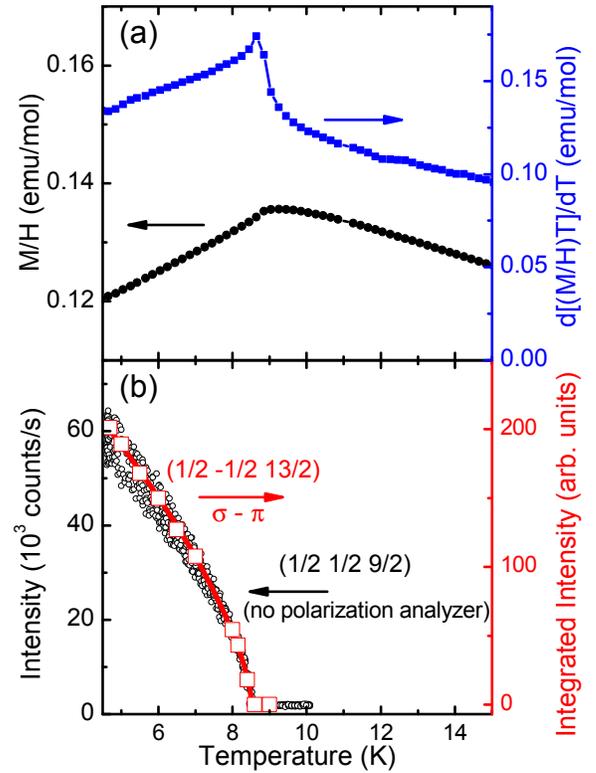}\\
\caption{(Color online) (a) $M/H$ and its temperature derivative
for GdBiPt. (b) The magnetic intensity measured while scanning
temperature at the maximum of the
($\frac{1}{2}\,\frac{1}{2}\,\frac{9}{2}$) diffraction peak
without a polarization analyzer (open small circles) and the
integrated intensity of the
($\frac{1}{2}$~-$\frac{1}{2}$~$\frac{13}{2}$) diffraction peak
measured at selected temperatures using the polarization analyzer
(open large squares).  The solid line is a power law fit to the
integrated intensity data as described in the text.} \label{fig3}
\end{figure}

The energy scan Fig.~\ref{fig2}(b) was performed with the
diffractometer set at the magnetic peak position and is typical of
resonant magnetic scattering at the $L$ edges of rare-earth
compounds.\cite{JWK_2005} At the $L_2$ edge of rare-earth
elements, the resonance primarily involves electric dipole ($E$1)
transitions from the 2$p_\frac{1}{2}$ core level to the empty 5$d$
states, seen as the strong line just at, or slightly below the
maximum in the measured fluorescence intensity.  The weaker
feature below the $E$1 resonance in Fig.~\ref{fig2}(b) is likely
due to the electric quadrupole ($E$2) transition from the
2$p_\frac{1}{2}$ core level to the 4$f$ states that are pulled
below the Fermi energy because of the presence of the core hole 
in the resonance process.

The temperature dependence of the magnetic scattering, along with 
the corresponding magnetization measurements performed on a 
sample from the same batch using a Quantum Design Magnetic 
Properties Measurement System, are shown in Fig.~\ref{fig3}.  The 
magnetic order parameter was measured at the 
($\frac{1}{2}\,\frac{1}{2}\,\frac{9}{2}$) peak position as the 
sample temperature was increased during a temperature scan in the 
absence of the polarization analyzer.  These data were 
supplemented by measurements of the integrated intensity of the 
($\frac{1}{2}$~-$\frac{1}{2}$~$\frac{13}{2}$) magnetic Bragg peak 
at selected temperatures and with polarization analysis. The line 
in Fig.~\ref{fig3}(b) describes a fit to the integrated intensity 
data using a power law of the form 
$I\,\sim\,(1-\frac{T}{T_N})^{2\beta}$ yielding 
$T_N$~=~8.52$\pm$0.05\,K and $\beta$~=~0.33$\pm$0.02. The close 
proximity of $T_N$ determined from our scattering measurements 
and the peak in d[($M/H$)T]\,/\,d$T$ 
(Ref.\,\onlinecite{Fisher_1962}) at $T$~=~8.6\,K, again confirms 
the magnetic origin of the Bragg scattering with a propagation 
vector of 
$\textbf{q}_m$~=~($\frac{1}{2}\,\frac{1}{2}\,\frac{1}{2}$).  
Systematic $M$ versus $H$ measurements (not shown) demonstrate in 
addition that no spontaneous ferromagnetic moment is present. 

\begin{figure}
\centering\includegraphics[width=0.88\linewidth]{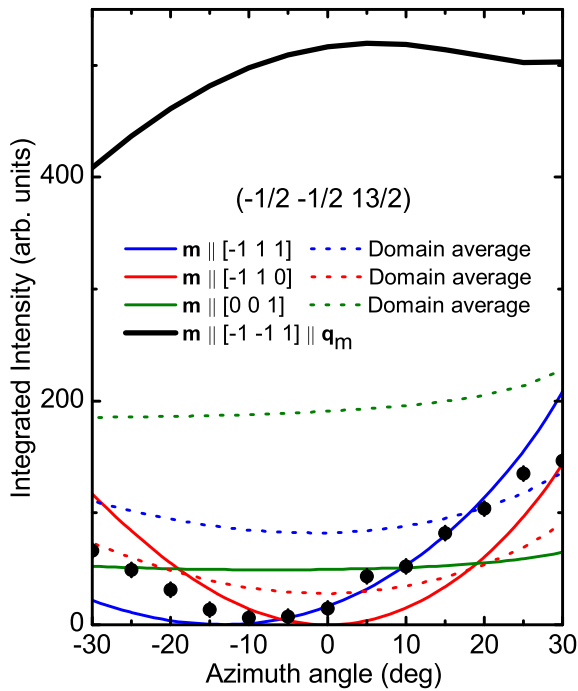}\\
\caption{(Color online) Integrated intensity in azimuth scans
through the (-$\frac{1}{2}$~-$\frac{1}{2}$~$\frac{13}{2}$)
magnetic Bragg peak. Measured data are depicted by full black
circles.  Full and dashed lines represent calculations for 
selected magnetic moment directions in a single magnetic domain, 
and for intensity from equally populated domains averaged over 
the three possible symmetry-equivalent magnetic moment 
orientations, respectively.} \label{fig4}
\end{figure}

Having established the nature of the magnetic ordering in GdBiPt, 
we now describe our attempt to determine the direction of the 
ordered magnetic moment.  The angular dependence of the resonant 
magnetic intensity $I(\psi)$ for the incident $\sigma$-polarized 
beam depends upon the component of the magnetic moment along the 
scattered beam direction and can be written as 
$I(\psi)_{(\textbf{Q},\,\alpha,\,\beta)}~=~C~[{\widehat{\textbf{m}}\cdot\widehat{\textbf{k}^\prime}(\psi)_{(\textbf{Q})}}]^2~A(\psi)_{(\textbf{Q},\,\alpha,\,\beta)}$ 
where C is an overall scale factor that accounts for the resonant 
scattering matrix element and incident beam intensity, 
$\widehat{\textbf{m}}$ and $\widehat{\textbf{k}^\prime}$ 
represent the magnetic moment and scattered beam directions, 
respectively, and $A$ accounts for the absorption 
correction.\cite{Detlefs_1997}  The sample geometry required 
off-specular scattering measurements of the magnetic peaks.  That 
is, the angle $\alpha$ of the incident beam $\textbf{k}$ with 
respect to the sample surface is different from the angle $\beta$ 
of the outgoing beam $\textbf{k}^\prime$ with respect to the 
sample surface.\cite{You_1999}  For the azimuth angle $\psi$ 
scans shown in Fig.~\ref{fig4}, the diffractometer was set at the 
position of the magnetic Bragg peak and the crystal was rotated 
around the scattering vector 
$\textbf{Q}\,=\,\textbf{k}^\prime-\textbf{k}$ thereby rotating 
$\widehat{\textbf{k}^\prime}$ with respect to 
$\widehat{\textbf{m}}$ while leaving $\textbf{Q}$ fixed. This 
yields an azimuth dependence of the intensity which is specific 
to a given magnetic moment direction.  Note, that the absorption 
correction $A$ also depends on the azimuth angle $\psi$.

For a cubic lattice, the determination of the ordered moment
direction is significantly complicated by the presence of domains
that arise from symmetry-equivalent magnetic propagation vectors 
and moment directions.  For the observed magnetic Bragg peaks at 
($H\,K\,L$) with $H$, $K$, and $L$ half integers, four 
symmetry-equivalent $\{\frac{1}{2}\,\frac{1}{2}\,\frac{1}{2}\}$ 
propagation vectors exist: 
($\frac{1}{2}$~$\frac{1}{2}$~$\frac{1}{2}$), 
(-$\frac{1}{2}$~-$\frac{1}{2}$~$\frac{1}{2}$), 
($\frac{1}{2}$~-$\frac{1}{2}$~-$\frac{1}{2}$), and 
(-$\frac{1}{2}$~$\frac{1}{2}$~-$\frac{1}{2}$).  Fortunately, for 
GdBiPt in the cubic space group $F\,\overline{4}\,3\,m$, only one 
propagation vector contributes to a particular magnetic 
reflection [e.\,g. the magnetic Bragg peak 
(-$\frac{1}{2}$\,-$\frac{1}{2}$\,$\frac{13}{2}$) is generated by 
the propagation vector 
$\textbf{q}_m$~=~(-$\frac{1}{2}$\,-$\frac{1}{2}$\,$\frac{1}{2}$) 
from the (0\,0\,6) zone center].  The measured data for the 
(-$\frac{1}{2}$\,-$\frac{1}{2}$\,$\frac{13}{2}$) magnetic Bragg 
peak show two important features in the azimuth scan presented in 
Fig.~\ref{fig4}: a distinct minimum with almost no intensity 
close to $\psi$~=~0, and an increase in intensity by more than an 
order of magnitude as $\psi$ is varied by $\pm30^\circ$.  Both 
features are in strong contrast to the expected $\psi$ dependence 
of the intensity for magnetic moments parallel to the propagation 
vector 
$\textbf{q}_m$~=~(-$\frac{1}{2}$\,-$\frac{1}{2}$\,$\frac{1}{2}$) 
as illustrated in Fig.~\ref{fig4} by the bold black line with a 
maximum close to $\psi$~=~0.  Therefore, we can exclude that the 
moments are parallel to the propagation vector in GdBiPt.

In Fig.~\ref{fig4}, calculated curves are also shown for other
moment directions.  Unfortunately, for each of the depicted moment
directions, three different symmetry-equivalent orientations can
occur yielding three magnetic domains.  The dashed lines in
Fig.~\ref{fig4} represent the calculated $\psi$ dependence of the
intensity if we include all such domains for a given moment
direction with equal population. We again find poor agreement
between the domain-averaged calculations for moments along the set
of \{1 1 1\}, \{1 1 0\} and \{0 0 1\} directions. However,
calculations assuming the presence of only a single domain within
the probed volume, with one specified moment direction (either 
[-1 1 1] or [-1 1 0] for the 
(-$\frac{1}{2}$\,-$\frac{1}{2}$\,$\frac{13}{2}$) Bragg peak in 
Fig.~\ref{fig4}) come much closer to describing the measured 
data. This behavior clearly indicates that the magnetic domains 
are large; smaller than the footprint of the incident beam on the 
sample (approximately 0.5$\times$0.5\,mm$^2$), but of the same 
order of magnitude. Similar large magnetic domains have been 
noted in previous XRMS work on GdNi$_2$Ge$_2$ as 
well.\cite{JWK_2005b} Nevertheless, a unique determination of the 
moment direction is not possible based on the available data. A 
more precise determination of the moment direction may be 
possible from measurements with much smaller incident beam 
dimensions and/or control of domain populations.\cite{JWK_2005b}

Summarising the experimental results, below $T_N$~=~8.5\,K the 
magnetic Gd moments order in a commensurate antiferromagnetic 
structure in GdBiPt that can be described as doubling the cubic 
unit cell along the diagonal [1\,1\,1] direction, so that 
alternating ferromagnetic (1\,1\,1) planes of Gd are 
antiferromagnetically coupled along the [1\,1\,1] direction. The 
moments are not aligned parallel to this diagonal [1\,1\,1] 
direction.

In contrast to GdBiPt, CeBiPt is an antiferromagnet characterized 
by a propagation vector $\textbf{q}_m$~=~(1\,0\,0) and the 
ordered moments are collinear with the propagation vector along 
[1\,0\,0],\cite{Wosnitza_2006} but with a reduced moment that 
may, in part, be attributed to crystalline electric field (CEF) 
effects.\cite{Goll_2007} Unfortunately, XRMS measurements do not 
allow a direct extraction of the ordered moment in GdBiPt, but 
earlier specific heat measurements\cite{Canfield_1991} estimated 
an entropy of $\sim$0.8\,$\mathcal{R}$\,ln\,8 associated with the 
magnetic transition close to the value expected for full moment 
ordering without CEF effects.  The entropy associated with the 
corresponding magnetic transitions for the Nd, Tb and Dy 
compounds were considerably less than $\mathcal{R}$\,ln(2$J$+1) 
expected for the full Hund's rule $J$ multiplet, indicating the 
importance of CEF effects in these compounds. The magnetic 
structures for $R$~= Nd, Sm, Tb, Dy, Ho, Er, Tm, and Yb, have not 
yet been identified by neutron or XRMS measurements and such 
measurments are planned.

Finally, we comment on our results in light of the proposal that 
GdBiPt may be an AFTI candidate.\cite{Mong_2010}  The AFTI state 
may be derived from either magnetic ordering in a pre-existing 
strong TI (model A in Ref.\,\onlinecite{Mong_2010}) or, 
alternatively, for specific antiferromagnetic ordering schemes 
that induce spin-orbit coupling in the system (model B in 
Ref.\,\onlinecite{Mong_2010}). Given previous ARPES 
measurements\cite{Liu_2011} above $T_N$, which do not find direct 
evidence for band inversion in GdBiPt, it seems unlikely that 
GdBiPt is itself a strong TI.  However, the magnetic structure 
determined here is consistent with the alternative model B 
presented by Mong \emph{et al.}\cite{Mong_2010}   The doubling 
along the cubic diagonal direction represents the broken lattice 
translational symmetry (by order of two) and the ordering of each 
magnetic moment breaks the time-reversal symmetry, however, the 
product of both symmetry operations is conserved for the 
determined magnetic order.  

We acknowledge valuable discussions with A.~Kaminski, 
J.\,E.~Moore, R.\,S.\,K.~Mong, P.\,J.~Ryan, and J.\,C.~Lang. This 
work was supported by the Division of Materials Sciences and 
Engineering, Office of Basic Energy Sciences, U.\,S. Department 
of Energy. Ames Laboratory is operated for the U.\,S. Department 
of Energy by Iowa State University under Contract No. 
DE-AC02-07CH11358. Use of the Advanced Photon Source was 
supported by the U.\,S.~DOE under Contract No. DE-AC02-06CH11357.

\bibliographystyle{apsrev}
\bibliography{gdbipt}

\end{document}